\batchmode
\makeatletter
\def\input@path{{C:/Arbeiten/}}
\makeatother
\documentclass[english]{article}
\usepackage[T1]{fontenc}
\usepackage[latin9]{inputenc}
\usepackage{color}
\usepackage{amsmath}
\usepackage{amssymb}
\usepackage{babel}
\begin{document}
\title{Null Foliations of Spacetime and the Geometry of Black Hole Horizons}
\author{Albert Huber\thanks{ahuber@tph.itp.tuwien.ac.at}}
\date{{\footnotesize{}Institut für theoretische Physik, Technische Universität
Wien, Wiedner Hauptstr. 8-10, A-1040 Wien, AUSTRIA}}
\maketitle
\begin{abstract}
In this work, a method for constructing null foliations of spacetime
is presented. This method is used to specify equivalence classes of
null generators, whose representatives can be associated lightlike
co-normals that are locally affine geodesic and thus locally orthogonal
to embedded null hypersurfaces of spacetime. The main benefit of the
proposed procedure is the fact that it is less geometrically restrictive
than the traditional dual-null approaches to general relativity, but
nevertheless allows for the conclusion that spacetimes can be foliated
by suitable pairs of normalized null geodesic vector fields. This
is demonstrated by the example of different black hole spacetimes,
that is, by members of the Kerr-Newman family, according to which
a said foliation and an associated equivalence class of null generators
are explicitly constructed.
\end{abstract}
\textit{\footnotesize{}Key words: general relatvity, spacetime, lightlike
foliations}{\footnotesize\par}

\section*{Introduction}

The problem of how to construct globally well-behaved null foliations
of spacetime (NFS) or double null foliations of spacetime (DNFS) is
a nontrivial issue in various regards.

First of all, in order to be able to carry out such a construction,
one generally is faced with the problem that lightlike hypersurfaces,
contrary to their spacelike and timelike counterparts, are non-Riemannian
submanifolds of spacetime, which are equipped with a degenerate metric
and a non-unique Levi-Civita connection. As a consequence, these hypersurfaces
possess geometric properties inherently different from other hypersurfaces
of spacetime and also from the Lorentzian manifold in which they are
embedded. In order to nevertheless find a natural starting point for
the construction of NFS or DNFS from a 4-geometric point of view,
it has become a customary procedure to characterize these hypersurfaces
indirectly by providing a 2+2-decomposition of spacetime and an associated
foliation of its manifold in spacelike 2-surfaces. However, while
this step certainly opens the door for constructing NFS or DNFS from
the induced geometries of their local slices, it usually turns out
that - by avoiding to work directly with the induced null geometry
of the hypersurface - it becomes difficult to retrospectively distinguish
intrinsically defined quantities of embedded null hypersurfaces from
others that conversely need to be extended off these hypersurfaces.

Besides this problem, there is a specific geometric subtlety associated
with the construction of NFS and DNFS, namely the fact that the generators
of lightlike hypersurfaces are generically not only orthogonal, but
also tangential to the submanifolds they generate. This characteristic
of null hypersurfaces makes it impossible to get off these hypersurfaces
by simply providing a null geodetic extension of their generators,
which implies that NFS can not simply be obtained by Lie transporting
some lightlike initial hypersurface along the flow of a generating
vector field. Instead, such a null geodesic extension must be performed
in practice - contrary to the standard non-null cases - with respect
to the alternate, non-tangential null normal of the said initial hypersurface,
whose flow, however, preserves its intrinsic geometric structure only
locally, if at all. 

In addition, there occurs the difficulty that a maximal extension
of the generator of an initial hypersurface along its co-normal generally
leads to caustics and thus to infinite values for both convergence
and shear of a null congruence of curves formed by the extended generator.
This makes it difficult to find handsome evolution equations and to
formulate a well-defined (characteristic) initial value problem within
the 2+2-framework of general relativity; a problem that has been tackled
from various different sides in the literature \cite{brady1996covariant,d1980covariant,friedrich1981asymptotic,friedrich1983characteristic,hayward1994general,isaacson1968harmonic,sachs1962characteristic,stewart1986characteristic,stewart1982numerical}. 

Next, there is the problem that the null gradient vector fields, which
lead to NFS or DNFS and which are well-defined from a null geometric
point of view, often are very difficult to find in practice. Especially
in the case of DNFS, whose constructions require that two independent
hypersurface forming null gradient vector fields and two associated
closed 2-forms are specified \cite{hayward1993dual}, it may turn
out to be a tricky endeavor to find gradients of scalar functions
that are real valued and additionally globally regular and therefore
compatible with the standard null geometric framework used in general
relativity. 

Ultimately, there remains the problem of providing an approach to
NFS or DNFS that retains the complete coordinate freedom of the underlying
theory. While a particular advantage of the considerations of \cite{hayward1993dual}
is that the respective 2+2-framework is both covariant and coordinate
independent, other approaches focussing on the construction of NFS
primarily concern the issue of developing well-defined null Gaussian
coordinates \cite{friedrich1999rigidity,moncrief1983symmetries} and
therefore do not attempt to give a coordinate independent description
of the problem in general. In turn, this makes it difficult to treat
- on the basis of these approaches - null geometric problems in a
universally conclusive way, which is further complicated by the fact
that the development of Gaussian null coordinates generally works
only locally.

In response to these difficulties, an alternative idea for providing
NFS is presented, which is based on the 2+2-framework of general relativity
and the standard construction method for DNFS. The said idea, which
is outlined in the very first section, is to consider a congruence
of null curves formed by a local null geodetic generator and to combine
this vector field into a whole equivalence class of such generators.
This is done by considering a continuing sequence of null rescalings,
which change the form of the respective null parameter in such a way
that the said null parameter can no longer be reasonably referred
to a given initial lightlike hypersurface. This leads to an equivalence
class of local null generators which all provide a NFS and whose existence
indirectly implies the fact that a given spacetime is additionally
foliated by another collection of lightlike hypersurfaces and thus
exhibits a DNFS. 

To demonstrate this, the construction of both NFS and DNFS is carried
out explicitly in the second section of the present work for the cases
of Schwarzschild spacetime in Kruskal-Szekeres and for Kerr and Kerr-Newman
spacetime in Kerr and in Hayward coordinates, respectively. 

\section*{Null Foliations of Spacetime and the 2+2-Framework of General Relativity}

Providing a foliation of spacetime may be viewed as one of the main
prerequisites for formulating a feasible theory of gravitation. 

This can be seen by the example of spacelike foliations, which form
the basis for the formulation of a well-defined Cauchy problem and
initial value formulation of the theory \cite{chrusciel2012einstein},
for its Hamiltonian description \cite{arnowitt1960canonical,arnowitt2008republication},
for a definition of quasilocal energy of the gravitational field \cite{arnowitt1959dynamical}
and for \textcolor{black}{the consistent canonical quantization of
Einstein-Hilbert gravity within the framework of canonical loop quantum
gravity \cite{ashtekar1991lectures,rovelli2007quantum,thiemann2007modern}.}

Additionally, it can be seen by the case of combined spacelike and
timelike foliations, whose construction is in general less clear,
but whose existence forms the basis for a definition of quasilocal
gravitational energy in general relativity, namely Brown's and York's
infamous generalization of the ADM mass \cite{brown1993quasilocal}. 

Finally, it can also be seen by the cases of NFS and DNFS. As already
emphasized before, these types of foliations represent a necessary
prerequisite for the characteristic initial value problem of general
relativity \cite{d1980covariant,friedrich1983characteristic,stewart1986characteristic}.
Aside from that, they have served - in a manner similar to spacelike
or spacelike and timelike foliations of spacetime - as a starting
point for providing different definitions of quasilocal gravitational
energy in general relativity, given by Hawking \cite{hawking1968gravitational,hawking1996gravitational}
and by Hayward \cite{hayward1994quasilocal}, respectively. Furthermore,
the consideration of DNFS has turned out to be a relevant factor in
generalizing the laws of black hole mechanics, i.e. in formulating
these laws also for dynamical black holes with the aid of the dual
null formalism \cite{hayward1994general}.

Focussing now specifically on NFS and DNFS, it ought to be clarified
that the respective constructions of these two different types of
foliations - despite of their related geometric output - generally
proceed totally differently from one another and typically start from
different theoretical assumptions. The only real similarity between
these approaches is the fact that they both rely on a 2+2-decomposition
of spacetime. Therefore, a reasonable step is to consider a spacetime
\textcolor{black}{$(M,g)$ with metric $g_{ab}$, which can be decomposed
in the }2+2-\textcolor{black}{form 
\[
g_{ab}=-2\ell_{(a}n_{b)}+2m_{(a}\bar{m}_{b)}
\]
}with respect to the quadruple of lightlike directions $(-n_{a},\,-\ell_{a},\,m_{a},\,\bar{m}_{b})$. 

Given such a decomposition of the spacetime metric, a NFS can be defined
as a collection of embedded null hypersurfaces. A precise explanation
of this statement can be given with respect to an open subset $\mathcal{O}$
of a spacetime $(M,g)$ and a three-dimensional lightlike submanifold
$\mathcal{H}$, defined in such a way that $\mathcal{H}\subset\mathcal{O}$.
Under the assumption that there exist local lightcone structures in
$\mathcal{O}$ and that $\mathcal{H}$, as a subset of $\mathcal{O}$,
can be viewed as a particular null hypersurface, which is generally
referred to as a so-called level set $\mathcal{H}=\mathcal{H}_{\sigma}$,
a foliation of a spacetime $(M,g)$ \textcolor{black}{is defined as
a collection of sections $\{\mathcal{H}_{\sigma}\}$ which vary smoothly
in $\sigma\in\mathbb{R}$ such that
\[
M=\underset{\sigma}{\bigcup}\mathcal{H}_{\sigma}.
\]
}Any given null hypersurface $\mathcal{H}$ contained in $(M,g)$
is therefore assumed to be labelled by a smooth parameter function
$\sigma$, which, of course, has to be constant along $\mathcal{H}$.
Thus it can be viewed as a particular representative of the set \textcolor{black}{$\{\mathcal{H}_{\sigma}\}$
that can be identified as an ordering prescription for all lightlike
hypersurfaces }of a spacetime $(M,g)$\textcolor{black}{.}

Considering the tangent vector field $\ell^{a}\in T(\mathcal{H})$
and its associated co-normal $n^{a}\in T(M)$, the consequence is
that there should hold 

\textcolor{black}{
\[
\ell_{a}\ell^{a}=0,\;\nabla_{[a}\ell_{b]}=0,\;\ell_{a}n^{a}=-1,
\]
}where the validity of the first two conditions of lightlikeness and
hypersurface orthogonality is non-optional, the validity of the third
condition of normalization, in return, is ususally introduced just
for reasons of simplicity. 

The respective conditions entail the instance that the integral curves
generated by $\ell_{a}$ have to be orthogonal to $\mathcal{H}$,
which is tantamount to the fact that $\ell_{a}$ has to fulfill the
Frobenius theorem\footnote{In fact, this is clear, as $\ell_{[a}\nabla_{b}\ell_{c]}=0\;\Longleftrightarrow\;^{*}\{\nabla_{[a}\ell{}_{b]}\}\ell^{a}=0$. }
\[
\ell_{[a}\nabla_{b}\ell_{c]}=0.
\]
A co-vector field $\ell_{a}\in T^{*}(\mathcal{H})$, which fulfills
the first two of the listed conditions, is called a lightlike generator
of a given null hypersurface, in the given case of $\mathcal{H}$. 

Since $\mathcal{H}$ is lightlike, its line bundle must be generated
by a lightlike co-vector field $-d\sigma_{a}=-\nabla_{a}\sigma$ fulfilling
the Eikonal equation $g^{ab}d\sigma_{a}d\sigma_{b}=0$, which, in
turn, requires $\ell_{a}$ and additionally $\ell^{a}$ to be affine
geodesic. Regarding the spin-coefficient formalism, this means that
there must hold $\varepsilon\text{+\ensuremath{\bar{\varepsilon}}}=\kappa=\tau-\bar{\alpha}-\beta=0$
and $\rho=\bar{\rho}$ with respect to the chosen null co-frame $(-n_{a},\,-\ell_{a},\,m_{a},\,\bar{m}_{b})$.
In consequence, the co-vector field $\ell_{a}\in T^{*}(\mathcal{H})$
ought to be given by the gradient of a scalar function, sometimes
called the optical function, which coincides with the previously introduced
'label parameter' $\sigma\in\mathbb{R}$ such that $\ell_{a}:=-d\sigma_{a}$.
Indeed, this is implied by the fact that in Lorentzian geometry no
torsion is present. 

As a further consequence of the listed conditions, one can determine
w.l.o.g. an associated, preferably \textcolor{black}{non-vanishing,
smooth, lightlike} vector field

\textcolor{black}{
\[
\ell^{a}=-g^{ab}d\sigma_{b},
\]
}whereas an important point is now that there is still a certain lattitude
in choosing a generator $\ell_{a}\in T^{*}(\mathcal{H})$ and its
longitudinal co-direction $n^{a}\in T(M)$. To be more precise, the
acquired properties of the null generator remain unchanged after a
multiplication with an arbitrary function $\chi=\chi(\sigma)$, by
which, as a consequence, one may equally obtain an alternative generator
$\ell'_{a}:=-\chi(\sigma)d\sigma_{a}$ of $\mathcal{H}$. This allows
one to conclude that there exists in fact a whole equivalence class
of generators of $\mathcal{H}$, which shall be denoted by $[\ell]$.

Looking then at the fact that spacetime is decomposed in spacelike
2-surfaces, one may regard a surface $\Delta\subset\mathcal{H}$ contained
in the respective spacelike foliation of $(M,g)$ and fix compatible
transversal lightlike directions $m_{a}\in T^{*}(\Delta),\;\bar{m}_{a}\in T^{*}(\Delta)$.
A foliation of $\mathcal{H}$ is obtained straightforwardly by Lie
transporting the generators of $\Delta$ along of the flow of the
null generator $\ell^{a}\in T(\mathcal{H})$. Although the section
$\Delta$ may possess a priori an arbitrary topological structure,
it may be chosen, if possible, to be homeomorphic to the two-dimensional
sphere $\mathbb{S}_{2}$ for convenience.

With this input, choosing local coordinates $x^{a}=(\sigma,\bar{\sigma},x^{2},x^{3})$
such that $\ell^{a}=\partial_{\bar{\sigma}}^{a}-L^{a}$, whereas $L^{a}\in T(\Delta)$
is a spacelike shift vector field, the line element of the spacetime
can be written down in the form

\[
ds^{2}=-2d\sigma d\bar{\sigma}+2L_{B}dx^{B}d\sigma+q_{BC}dx^{B}dx^{C}+L_{B}L^{B}d\sigma^{2},
\]
whereby $B,C=2,3$ and $q_{BC}:=2m_{(B}\bar{m}_{C)}$. By performing
a coordinate transformation $\bar{\sigma}=\sigma-\rho$, this line
element can be rewritten w.l.o.g. in the form 
\[
ds^{2}=-\phi d\sigma^{2}+2d\sigma d\rho+2L_{B}dx^{B}d\sigma+q_{BC}dx^{B}dx^{C},
\]
which shows that the given line element coincides with the null Gaussian
coordinate system of Moncrief and Isenberg \cite{moncrief1983symmetries}
and thus, in the further course, with that of Friedrich, Racz and
Wald \cite{friedrich1999rigidity} in addition. The resulting null
coordinates are adapted to characteristic null hypersurfaces of spacetime
and therefore of great interest in various regards in general relativity.

Turning from this to the subject of DNFS, it shall be pointed out
first that a DNFS \textcolor{black}{is defined }in contrast\textcolor{black}{{}
-} provided the fact that the same basic setting is given as before
-\textcolor{black}{{} as a collection of sections }$\{\{\mathcal{H}_{\sigma}\},\{\mathcal{\bar{H}}_{\bar{\sigma}}\}\}$\textcolor{black}{{}
varying smoothly in the parameters $\sigma,\bar{\sigma}\in\mathbb{R}$
such that} there holds

\[
M=\underset{\sigma}{\bigcup}\mathcal{H}_{\sigma}=\underset{\bar{\sigma}}{\bigcup}\mathcal{\bar{H}}_{\bar{\sigma}}.
\]
Contrary to NFS, DNFS are generally characterized by a pair orthogonal
connecting vectors\textcolor{black}{{} of some spacelike 2-surface $\Delta$,
which is completed to a pseudo-orthonormal coordinate frame $(\sigma^{a},\bar{\sigma}^{a},e_{2}^{a},e_{3}^{a})$
consecutively. The directions $\sigma^{a}:=\partial_{\sigma}^{a}$
and $\bar{\sigma}^{a}:=\partial_{\bar{\sigma}}^{a}$, called the evolution
vector fields, are defined in such a way that the given spacetime
basis can be Lie propagated, i.e. $\phi_{\sigma}(\sigma^{a},\bar{\sigma}^{a},e_{2}^{a},e_{3}^{a})=\phi_{\bar{\sigma}}(\sigma^{a},\bar{\sigma}^{a},e_{2}^{a},e_{3}^{a})=0$,
guaranteeing that $(\sigma^{a},\bar{\sigma}^{a},e_{2}^{a},e_{3}^{a})$
is holonomic. This concrete choice does not only allow a development
of spacetime in a neighborhood $\tilde{\Delta}$ of $\Delta$, but
also results in a formation of two hypersurfaces $\mathcal{H}$ and
$\bar{\mathcal{H}}$ by the procedure of applying the pair of flows
$(\phi_{\sigma},\phi_{\bar{\sigma}})$ to vectors tangent to $\Delta$,
meaning actually $\mathcal{H}_{\sigma}=\phi_{\sigma}(\Delta)$ and
$\mathcal{\bar{H}}_{\bar{\sigma}}=\phi_{\bar{\sigma}}(\Delta)$. This
gives a foliation of the whole spacetime into pairs of null surfaces,
ergo a }DNFS\textcolor{black}{.}

\textcolor{black}{Each fixed 2-surface $\Delta$ then can be thought
of as the apex of $\sigma=const.$-surfaces and $\bar{\sigma}=const.$-surfaces,
which is a prerequisite for the considering of a dual-null geodesic
frame. }As a result, there ought to be affine geodesic vector fields
$\ell^{a}\in T(\mathcal{H})$ and $n^{a}\in T(\bar{\mathcal{H}})$
associated with the pair of hypersurfaces $\mathcal{H},\,\bar{\mathcal{H}}\subset M$\textcolor{black}{,
which need to fulfill the conditions
\[
\ell_{a}\ell^{a}=n_{a}n^{a}=0,\;\nabla_{[a}\ell_{b]}=\nabla_{[a}n_{b]}=0,\;\ell_{a}n^{a}=-e^{m}.
\]
It is }a nontrivial task to obtain two closed 2-forms and thus to
sastisfy the\textcolor{black}{{} listed conditions on a generic spacetime.
}In terms of spin-coefficients, the said conditions encode the relations
$\varepsilon\text{+\ensuremath{\bar{\varepsilon}}}=\kappa=\tau-\bar{\alpha}-\beta=0$,
$\rho=\bar{\rho}$ and $\varepsilon^{'}\text{+\ensuremath{\bar{\varepsilon}}}^{'}=\kappa^{'}=\tau^{'}-\bar{\alpha}^{'}-\beta^{'}=0$
and $\rho^{'}=\bar{\rho}^{'}$. Thus, looking at the complexity of
these conditions, it immediately becomes clear why the normalization
condition generally has to be dropped\textcolor{black}{{} in the dual-null
framework. }

\textcolor{black}{If the listed conditions nevertheless can be satisfied,
both of the vector fields orthogonal to $\Delta\subset M$ must be}
generators and thus both must be given by $\ell_{a}:=-d\sigma_{a}$
and $n_{a}:=-d\bar{\sigma}_{a}$ with respect to the pair of optical
functions \textcolor{black}{$\sigma,\bar{\sigma}\in\mathbb{R}$. As
a result, $n_{a}\in T^{*}(\bar{\mathcal{H}})$ }now also fulfills
\[
n_{[a}\nabla_{b}n_{c]}=0
\]
\textcolor{black}{along each three dimensional lightlike co-hypersurface
$\bar{\mathcal{H}}\in\{\mathcal{\bar{H}}_{\bar{\sigma}}\}$.} In local
coordinates $x^{a}=(\sigma,\bar{\sigma},x^{2},x^{3})$ one finds that
the generating vector fields must possess the form

\[
\ell^{a}=e^{m}(\partial_{\bar{\sigma}}^{a}-L^{a})
\]
and
\[
n^{a}=e^{m}(\partial_{\sigma}^{a}-N^{a})
\]
respectively. In this context, the largely undetermined function $e^{m}$
with $m=m(x)$ represents an analogon to the lapse function in the
dual-null case and $L^{a}\in T(\mathcal{H})$ and $N^{a}\in T(\bar{\mathcal{H}})$
are shift vector fields, sometimes referred to as equivariant vector
fields. This form immediately leads to the following decomposition
of the line element of $(M,g)$

\[
ds^{2}=-2e^{-m}d\sigma d\bar{\sigma}+q_{AB}(dx^{A}+L^{A}d\sigma+N^{A}d\bar{\sigma})(dx^{B}+L^{B}d\sigma+N^{B}d\bar{\sigma}).
\]
Obviously, one re-obtains from this line element, for the special
case of $e^{m}=1$ and $N^{A}=0$, once more a null Gaussian coordinate
system. 

To proceed, by looking now closely at the addressed settings, one
may realize that one of the most essential differences between the
construction of NFS and DNFS is the fact that the two approaches are
based on the use of different null geodesic frames. While the construction
of NFS is generally based on the use of normalized null geodesic frames,
the construction of DNFS is based on the use of dual-null geodesic
frames instead. Therefore, there naturally arises the question if
and how it is possible to transite from one approach to the other. 

To answer this question, consider the above dual-null setting and
assume that the pair of co-vector fields $\ell_{a},n_{a}\in T^{*}(M)$
have been completed to a null co-tetrad $(-n_{a},\,-\ell_{a},\,m_{a},\,\bar{m}_{b})$.
It quickly becomes clear that the conditions 

\textcolor{black}{
\[
\ell_{a}\ell^{a}=0,\;\nabla_{[a}\ell_{b]}=0,\;\ell_{a}n^{a}=-1
\]
}easily can be met if one assumes $n_{a}\propto-e^{-m}d\bar{\sigma}_{a}$
in order to transition to a normalized null geodesic frame. \textcolor{black}{Thus,
satisfying the normalization condition $\ell_{a}n^{a}=-1$ comes at
the price of giving up the hypersurface orthogonality of one of the
generators. }

The\textcolor{black}{{} basic idea }in order to obtain a suitable normalized
null geodesic tetrad containing the generator of a NFS in \textcolor{black}{sets
of level surfaces $\{\mathcal{H}_{\sigma}\}$} \textcolor{black}{is
the following: Consider a lightlike hypersurface }$\mathcal{H}\equiv\mathcal{H}_{0}$\textcolor{black}{{}
embedded into a spacetime $(M,g)$. This hypersurface, which by assumption
is generated by a null vector field $\ell^{a}$, shall be given in
such a way that it is intersected by another lightlike hypersurface
$\bar{\mathcal{H}}$} in the apex $\Delta$, whose generator is non-tangential
to $\mathcal{H}$. In relation to this geometric setting, after\textcolor{black}{{}
introducing once more local coordinates }$x^{a}=(\sigma,\bar{\sigma},x^{2},x^{3})$,
it can be concluded that by choosing the generating co-vector field
as a scalar multiple of a gradient, i.e. by choosing it to be of the
form $\ell_{a}=-\chi(\sigma)d\sigma_{a}$, one clearly determines
an affine geodesic hypersurface orthogonal vector field whose 4-geometric
structure clearly is compatible with the stucture of the intrinsically
defined local null fields. Beyond that one knows that locally $n_{a}\arrowvert_{\mathcal{H}}=-d\bar{\sigma}_{a}$
must be valid in the case that $n^{a}$ is assumed to coincide locally
with the generator of $\bar{\mathcal{H}}$, which intersects \textcolor{black}{$\mathcal{H}$
in $\Delta$.} Accordingly, one can make the ansatz $\ell^{a}\arrowvert_{\mathcal{H}}=\partial_{\bar{\sigma}}^{a}-L_{0}^{a}$,
which directly implies that $\ell^{a}=g(\partial_{\bar{\sigma}}^{a}-L^{a})$,
whereas $g$ is some scalar function that can be chosen w.l.o.g. to
be of the form $g=\chi^{-1}e^{m}$ with $\chi\equiv e^{m_{0}}$, where
$m_{0}=m_{0}(\sigma)$ by definition shall apply in such a way that
also $m_{0}-m\vert_{\Delta}=const.$ applies. An obvious and consistent
choice for $n^{a}$ is then $n^{a}=e^{-m_{_{0}}}(\partial_{\sigma}^{a}-N^{a})$\textcolor{black}{.}

As a result of these local considerations, one is finally left with
the following collection of fields

\[
\ell_{a}=-e^{-m_{_{0}}}d\sigma_{a},\;\ell^{a}=e^{m-m_{0}}(\partial_{\bar{\sigma}}^{a}-L^{a});\;\;n_{a}=-e^{m_{0}-m}d\bar{\sigma}_{a},\;n^{a}=e^{m_{_{0}}}(\partial_{\sigma}^{a}-N^{a}),
\]
which satisfies all of the considered requirements.

Based on the fact that the given steps can be performed with respect
to any given fixed hypersurface $\mathcal{H}$ and then certainly
be repeated with respect to any other null hypersurface $\mathcal{H}'$,
which intersects the null hypersurface $\bar{\mathcal{H}}$ in another
apex $\Delta'$, one is thereby left with a NFS that is fully compatible
with the previously addressed dual-null framework and the general
construction of DNFS. 

Considering the class of null fields associated with $\mathcal{H}'$
, one finds that these fields once more must be of the form 
\[
\ell'_{a}=-e^{-m_{1}}d\sigma_{a},\;\ell'^{a}=e^{m-m_{1}}(\partial_{\bar{\sigma}}^{a}-L^{a});\;\;n'_{a}=-e^{m_{1}-m}d\bar{\sigma}_{a},\;n'^{a}=e^{m_{_{1}}}(\partial_{\sigma}^{a}-N^{a}),
\]
where $m_{1}=m_{1}(\sigma).$ This shows that $\ell'^{a}$ lies in
the equivalence class $[\ell]$ of $\ell^{a}$ and vice versa. Accordingly,
as the standard rescaling freedom of null normals allows one to reach
any fixed parameter value $\sigma_{i}$ of $\sigma$ and therefore
any given portion $\Delta_{i}$ lying either in the foliation of $\mathcal{H}$
or in the foliation of $\bar{\mathcal{H}}$, it appears that the information
that $(M,g)$ is additionally foliated by a set $\{\mathcal{\bar{H}}_{\bar{\sigma}}\}$
of $\bar{\sigma}=const.$-hypersurfaces now is encoded in the structure
of $[\ell]$. Hence, for the above choice of null fields,\textcolor{black}{{}
the }class $[\ell]$ naturally provides an\textcolor{black}{{} equivalence
class $[\dot{x}]$ of vector fields $\dot{x}^{a}$, which is defined
with respect to any given parameter value $e^{m_{_{i}}}$ of $e^{m}$
and which generates a geodesic congruence providing a foliation of
the regarded spacetime in lightlike hypersurfaces. }

In consequence, this method must lead to a similar decomposition of
the line element of $(M,g)$

\[
ds^{2}=-2e^{-m}d\sigma d\bar{\sigma}+q_{AB}(d\theta^{A}+L^{A}d\sigma+N^{A}d\bar{\sigma})(d\theta^{B}+L^{B}d\sigma+N^{B}d\bar{\sigma})
\]
as the dual-null framework. This is interesting insofar as that one
would expect that a null Gaussian coordinate system in the sense of
Moncrief and Isenberg would be - at least locally - much more perfectly
adapted to the existence of a NFS. However, this obviously does not
seem to be the case here; at least not at first sight. Only by looking
more closely one sees that the introduction of null Gaussian coordinates
can straightforwardly be achieved in the case that $N^{A}=0$ by transforming
the above null coordinates. To be more specific, the given coordinate
system can be transformed in that case into a null Gaussian coordinate
system by a transformation of the form $\bar{\sigma}=\bar{\sigma}(m)$,
which is chosen in such a way that $\bar{\sigma}=-e^{m}$. This results
in a line element which is again of the form

\[
ds^{2}=-\phi d\sigma^{2}+2d\sigma dm+2\beta_{B}dx^{B}d\sigma+q_{AB}dx^{A}dx^{B}
\]
with $\phi$, $\beta_{B}$ and $q_{AB}$ all being functions of $(\sigma,m,x^{2},x^{3})$.
Note that the given class of spacetimes contains the important Robinson-Trautmann
class of spacetimes \cite{robinson1962some} as a special case, which
is a class of spacetimes foliated by a hypersurface-orthogonal, shear-free
and expanding congruence of null curves. 

Following \cite{pawlowski2004spacetimes}, there occurs hereby an
interesting side aspect of the theory of NFS if $q_{AB}=q_{AB}(m,x^{2},x^{3})$.
This instance is based on the fact that a spacetime geometry whose
line element is of the present type necessarily belongs to a subclass
of the Robinson-Trautmann class known as the Kundt class of spacetimes,
which can be foliated by non-expanding horizons. This is due to the
fact that this very class of spacetimes is one for which there exists
a null congruence of curves which is produced by a null generator
that is both non-expanding and non-shearing. Therefore, it is a class
of spacetimes with topology $\mbox{\ensuremath{\mathbb{R}}}\times\mbox{\ensuremath{\mathbb{R}}}\times\mathbb{S}_{2}$,
which is foliated by null hypersurfaces on which locally the null
dominant energy condition is satisfied. 

As a direct result of this fact, it becomes clear that it must be
possible to associate to each so-called non-expanding horizon $\mathcal{H}$
in the null foliation of such a spacetime an equivalence class $[\ell]$
of vector fields, whose representatives ought to fulfill

\[
[\text{£}_{\ell},\underset{\leftarrow}{\nabla_{b}}]\ell^{a}=0,
\]
where $\text{£}_{\ell}$ denotes the Lie-derivative along $\ell^{a}\in[\ell]$
and $\underset{\leftarrow}{\nabla_{a}}$ the covariant derivative
on $\mathcal{H}$, respectively. This condition is tantamount to requiring
that there exists a rotational 1-form potential $\omega_{a}$ for
which there holds $\text{£}_{\ell}\omega_{a}=\iota_{a}^{c}L_{\ell}\omega_{c}=0,$
where $L_{\ell}$ denotes the Lie-derivative along $\ell^{a}\in[\ell]$
in $(M,g).$ Vector fields satisfying these conditions form what may
be referred to as the 'internal equivalence class' $[\ell]$ of generating
null vectors. This equivalence class has to be distinguished from
the previously addressed, homonymous 'external equivalence class'
$[\ell]$ encoding the existence of an additional NFS, from which
this internal class straightforwardly can be constructed in the second
place. Given such an internal equivalence class $[\ell]$, the resulting
pair $(\mathcal{H},[\ell])$ therefore defines locally a so-called
extremal weakly isolated horizon. Furthermore, in case that $[\ell]$
is so restrictive that even 
\[
[\text{£}_{\ell},\underset{\leftarrow}{\nabla_{b}}]q^{a}=0
\]
can be fulfilled with respect to all vectors $q^{a}\in T(\mathcal{H})$,
the pair $(\mathcal{H},[\ell])$ defines a so-called extremal isolated
horizon, which is embedded in a NFS. A horizon of this kind represents
a well-known generalization of the notion of a Killing horizon, which
plays an important role in gravitational physics. As it turns out
in the case of pp-wave spacetimes, which are the most prominent examples
of Kundt spacetimes, the only principal null direction of the geometry
coincides exactly with the Killing vector field of these geometries,
so that it can be concluded that these special types of spacetimes
can always be foliated by non-expanding Killing horizons. 

As a next step, it would also be interesting to know whether or not
a black hole spacetime can in principle be foliated by a collection
expanding null hypersurfaces containing a subcollection of non-expanding
Killing horizons, which would require to find a restricted null Gaussian
coordinate system, in respect to which both the Eikonal equation and
the scalar covariant wave equation can be solved at the same time
in regard to one and the same real-valued scalar field. 

In the case of stationary black hole spacetimes, the next chapter
will demonstrate that a corresponding foliation actually exists, showing
that the different geometric frameworks of the null and the dual null
approaches to black hole physics can reasonably be reconciled.

\section*{\textcolor{black}{Lightlike Foliations for stationary Black Hole
Spacetimes}}

\textcolor{black}{The line of arguments presented in the previous
section is now applied to two concrete examples: to Schwarzschild
spacetime in Kruskal-Szekeres coordinates and to Kerr-Newman spacetime
in Kerr and Hayward coordinates, respectively. In both cases, explicit
calculations are accommodated which lead to a class of generating
vector fields and thus to a foliation of both geometries in lightlike
hypersurfaces.}

\subsection*{\textcolor{black}{A lightlike Foliation of Schwarzschild Spacetime}}

\textcolor{black}{It is well-known that the Schwarzschild geometry
possesses an analytic continuation provided by a change from Schwarzschlild
to Kruskal-Szekeres coordinates. In these coordinates, the line element
of the spacetime takes the form}

\textcolor{black}{$$d s^{2} = -2 A d U dV + r^{2} d \Omega^{2},$$
where $r = r(UV)$ is implicitly given by $UV = (1 - \frac{r}{2M})e^{\frac{r}{2M}}$
and $A=A(r(UV))= \frac{16 M^{3}}{r}e^{-\frac{r}{2M}}$. According
to that setting, Schwarzschild spacetime decomposes into four autonomic
parts I - IV, i.e. into two asymptotically flat regions and two regions
containing a singularity at $r = 0$. The two lightlike hypersurfaces,
the $V=0$- and the $U=0$-surfaces, which divide the spacetime into
the four regions, constitute the horizon and the co-horizon and shall
be denoted as $\mathcal{H}$ and $\bar{\mathcal{H}}$. The intersection
hypersurface $\Delta = \mathcal{H} \cap \bar{\mathcal{H}}$ associated
with the value $r=2M$ of the implicit function $r = r(UV)$ has the
attributes that it is a bifurcation hypersurface on one hand and a
particular leaf of the individual foliations of $\mathcal{H}$ and
$\bar{\mathcal{H}}$ on the other hand. }An alternate 2-surface $\Delta'$
created by the intersection of a \textcolor{black}{$V=V_{0}=const.$-surface
$\mathcal{H}'$ with the fixed $U=0$-hypersurface $\mathcal{\bar{H}}$
shall furthermore be considered. }

\textcolor{black}{With that being given, both hypersurfaces $\mathcal{H}$
and $\bar{\mathcal{H}}$ can straightforwardly be identified as folia
of a NFS of Schwarzschild. This can be seen by setting $e^{-m}=A(r(UV))$
and by realizing that $L^{a}=N^{a}=0$. Comparing this with the considerations
of the previous section one can make the assignment $e^{m_{0}-m} = \frac{A(UV_{0})}{A(UV)}$.
One therefore sees that} \textcolor{black}{the generator of $\mathcal{H}'$
emanating from $\Delta'$, which is also a portion of the lightlike
hypersurface $\bar{\mathcal{H}}$, has to posses the structure $\ell^{a}\arrowvert_{\mathcal{H}'}=\partial_{V}^{a}$. }

\textcolor{black}{Repeating then the main steps of the previous section,
one immediately is left with the following locally defined collection
of fields}

\[
n^{a}=\frac{1}{A(V_{0}U)}\partial_{U}^{a},\;n_{a}=-\frac{A(UV)}{A(UV_{0})}dV_{a},\;\ell^{a}=\frac{A(UV_{0})}{A(UV)}\partial_{V}^{a},\;\ell_{a}=-A(UV_{0})dU_{a}.
\]
\textcolor{black}{According to that particular choice, the vector
field $\ell^{a}$ delivers a foliation in $\sigma=A(UV_{0})=const.$-hypersurfaces,
yielding finally the integral curves $\dot{x}^{a} = \frac{d x^{a}}{d \sigma}, $
whose equivalence class can be set up in such a way that it produces
a co-foliation in $\bar{\sigma}=A(U_{0}V)=const.$-hypersurfaces.
The resulting curves have to form}

\begin{align*}
\dot{U} & =0,\\
\dot{V} & =\frac{A(UV_{0})}{A(UV)},\\
\dot{\theta} & =0,\\
\dot{\phi} & =0.
\end{align*}
Obviously, a rescaling by a function $f(U)=\frac{A(UV_{1})}{A(UV_{0})}$
yields again a collection of vector fields
\[
n^{a}=\frac{1}{A(V_{1}U)}\partial_{U}^{a},\;n_{a}=-\frac{A(UV)}{A(UV_{1})}dV_{a},\;\ell^{a}=\frac{A(UV_{1})}{A(UV)}\partial_{V}^{a},\;\ell_{a}=-A(UV_{1})dU_{a}.
\]
belonging to the same equivalence class $[\ell],$ delivering the
same integral curves for a different initial data $A(UV_{1}).$ Since
this step can be repeated continuously with respect to any given fixed
value $V_{n}$ of $V$ associated with the fixed, but completely arbitrary
$V=const.$-hypersurface \textcolor{black}{$\bar{\mathcal{H}}$,}
$[\ell]$ creates a congruence of integral curves which foliates Schwarzschild
spacetime in the considered coordinates.

Completing in these coordinates $\ell^{a}$ to a tetrad field $(\ell^{a},\,n^{a},\,m^{a},\,\bar{m}^{a})$
of the form
\begin{align*}
\ell^{a} & =\frac{A(UV_{1})}{A(UV)}\partial_{V}^{a},\\
n^{a} & =\frac{1}{A(V_{1}U)}\partial_{U}^{a},\\
m^{a} & =\frac{1}{r}(\partial_{\theta}^{a}+\frac{i}{\sin\theta}\partial_{\phi}^{a}),\\
\bar{m}^{a} & =\frac{1}{r}(\partial_{\theta}^{a}-\frac{i}{\sin\theta}\partial_{\phi}^{a}),
\end{align*}
one sees that the resulting NFS contains a non-expanding horizon at
$U=0$ with a null normal that is locally a scalar multiple of the
Killing vector field $\xi^{a}=U\partial_{U}^{a}+V\partial_{V}^{a}$.
One therefore knows that this particular hypersurface is a Killing
horizon and therefore an isolated horizon as well. 

Finally, one can see by transforming to Eddington-Finkelstein coordinates
that the Schwarzschild geometry 
\[
ds^{2}=-(1-\frac{2M}{r})dv^{2}+2dvdr+r^{2}d\Omega^{2}
\]
allows one to define locally a null Gaussian coordinate system in
the sense of Monrief and Isenberg.

Therefore, it is in fact not difficult to realize that the normalized
null co-vector field $\ell_{a}=-dv_{a}$ solves indeed the Eikonal
equation in these coordinates and thereby provides a NFS of Schwarzschild
spacetime. The same holds true, in fact, for any associated null geodesic
co-vector field $\ell_{a}=-df_{a}$ (as long as $f=f(v)$) and in
particular for the specific null co-vector field that was constructed
above, which in the given coordinates takes the particular form $\ell_{a}=-c\cdot e^{\kappa(v-v_{0})}dv_{a}$,
where $c$, $\kappa$ and $v_{0}$ are all constants. Therefore both
vector fields are found to lie in the same equivalence class, which
shows that a direct transition from the dual null to the given null
geometric framework can be achieved in the present context by simply
introducing a specific, but finite sequence of coordinate transformations. 

\subsection*{\textcolor{black}{A lightlike Foliation of Kerr-Newman Spacetime}}

Given a NFS of Schwarzschild, the logical next step is to provide
a similar structure for Kerr and, at the same time, for Kerr-Newman
spacetime. To do so, at first sight, it appears to be reasonable to
consider coordinates that are regular on both Killing horizons. Thus,
one may consider Kerr coordinates, which fulfill precisely these requirements.
In these coordinates, the Kerr-Newman line element reads 
\[
ds^{2}=-(1-\frac{2Mr-e^{2}}{\Sigma})dv^{2}+2(dv-a\sin^{2}\theta d\phi)dr+\Sigma d\theta^{2}+
\]

\[
+\frac{\Pi\sin^{2}\theta}{\Sigma}d\phi^{2}-\frac{2(2Mr-e^{2})}{\Sigma}a\sin^{2}\theta dvd\phi
\]
where $\Sigma=r^{2}+a^{2}\cos^{2}\theta$, $\Pi=(r^{2}+a^{2})^{2}-\Delta a^{2}\sin^{2}\theta$
and $\Delta=r^{2}+a^{2}-2Mr+e^{2}$. The inverse metric can be read
off from 

\[
ds^{-2}=\frac{a^{2}\sin^{2}\theta}{\Sigma}\partial_{v}^{2}+\frac{2(r^{2}+a^{2})}{\Sigma}\partial_{v}\partial_{r}+\frac{\Delta}{\Sigma}\partial_{r}^{2}+\frac{2a}{\Sigma}\partial_{v}\partial_{\phi}
\]

\[
+\frac{2a}{\Sigma}\partial_{r}\partial_{\phi}+\frac{1}{\Sigma}\partial_{\theta}^{2}+\frac{1}{\sin^{2}\theta\Sigma}\partial_{v}^{2}.
\]
Indeed one immediately recovers Kerr as a special case from Kerr-Newman
by setting $e=0$. 

Given this setting, the question is once more how the conditions 

\textcolor{black}{
\[
\ell_{a}\ell^{a}=0,\;\nabla_{[a}\ell_{b]}=0,\;\ell_{a}n^{a}=-1
\]
}can be fulfilled. By performing a $2+2$-decomposition of the metric,
one finds immediately that none of the coordinate null vector fields
in these coordinates is generating except for the local case of $\Delta=0$,
i.e. for the pair of interior and exterior Killing horizons $\mathcal{H}^{\pm}$
of the black hole. 

Thus, one has to look instead at the geometric structure of the inverse
metric, in respect to which however one directly finds other meaningful
candidates, namely the pair of generators 
\[
\ell_{a}^{\pm}=dv_{a}+f_{\pm}dr_{a}\pm a\cos\theta d\theta_{a}
\]
and

\[
\ell^{\pm a}=(\frac{a^{2}\sin^{2}\theta}{\Sigma}+\frac{r^{2}+a^{2}}{\Sigma}f_{\pm})\partial_{v}^{a}+(\frac{r^{2}+a^{2}}{\Sigma}+\frac{\text{\ensuremath{\Delta}}}{\Sigma}f_{\pm})\partial_{r}^{a}+(1+f_{\pm})\frac{a}{\Sigma}\partial_{\phi}^{a}\pm\frac{a\cos\theta}{\Sigma}\partial_{\theta}^{a}
\]
respectively, according to which $f_{\pm}=-\frac{r^{2}+a^{2}}{\Delta}\pm\frac{\Lambda}{\Delta}$
and $\Lambda=((r^{2}+a^{2})^{2}-a^{2}\Delta)^{\frac{1}{2}}=(r^{4}+a^{2}r^{2}+a^{2}(2Mr-e^{2}))^{\frac{1}{2}}$. 

Given these fields, at first sight, it appears as if both generators
would become singular at $r=r_{\pm}=M\pm\sqrt{M^{2}-a^{2}-e^{2}}$.
However, if one uses the power series expansion $f_{\pm}=\frac{r^{2}+a^{2}}{\Delta}(-1\pm(1-\frac{1}{2}\frac{a^{2}\Delta}{(r^{2}+a^{2})^{2}}+...)$,
one sees that $f_{+}$ remains perfectly regular at $r=r_{\pm}$.
Thus, the co-vector field 
\[
\ell_{a}=dv_{a}+f_{+}dr_{a}+a\cos\theta d\theta_{a}
\]
reperesents a well-defined generator and surely provides a foliation
in $\sigma\equiv v+\int f_{+}dr+a\sin\theta=const.$-hypersurfaces.
Taking further into account that $\frac{1}{f_{+}}\vert_{r=r_{\pm}}=-\frac{2(r_{\pm}^{2}+a^{2})}{a^{2}}$,
one can make in connection thereto the following, perfectly appropriate
choice for its co-normal 
\[
n_{a}=-\frac{1}{f_{+}}(dv_{a}-a\sin^{2}\theta d\phi_{a}),\;n^{a}=-\frac{1}{f_{+}}\partial_{r}^{a}.
\]
Ultimately, this allows one to fulfill all of the imposed conditions,
which leads to integral curves\textcolor{black}{{} of the form}

\begin{align*}
\dot{v} & =\frac{a^{2}\sin^{2}\theta}{\Sigma}+\frac{r^{2}+a^{2}}{\Sigma}f_{+},\\
\dot{r} & =\frac{r^{2}+a^{2}}{\Sigma}+\frac{\text{\ensuremath{\Delta}}}{\Sigma}f_{+},\\
\dot{\theta} & =\frac{a\cos\theta}{\Sigma},\\
\dot{\phi} & =(1+f_{+})\frac{a}{\Sigma}.
\end{align*}
Using then the following Kerr-Schild decomposition of the line element
\[
ds^{2}=\frac{a^{2}\sin^{2}\theta}{\Sigma}dv^{2}+2(dv-a\sin^{2}\theta d\phi)dr+\Sigma d\theta^{2}-\frac{2(2Mr-e^{2})}{\Sigma}a\sin^{2}\theta dvd\phi+
\]
\[
+\frac{(r^{2}+a^{2})^{2}}{\Sigma}\sin^{2}\theta d\phi^{2}-\frac{\Delta}{\Sigma}(dv-a\sin^{2}\theta d\phi)^{2},
\]
one finds by introducing a new coordinate $v\rightarrow\sigma\equiv v+\int f_{+}dr+a\sin\theta$
that the Kerr-Newman line element can be written down in the form
\[
ds^{2}=\frac{a^{2}\sin^{2}\theta}{\Sigma}(d\sigma-f_{+}dr-a\cos\theta d\theta)^{2}-2(d\sigma-f_{+}dr-a\cos\theta d\theta+a\sin^{2}\theta d\phi)dr
\]
\[
+\Sigma d\theta^{2}+\frac{(r^{2}+a^{2})^{2}}{\Sigma}\sin^{2}\theta d\phi^{2}+\frac{2(2Mr-e^{2})}{\Sigma}a\sin^{2}\theta(d\sigma-f_{+}dr-a\cos\theta d\theta)d\phi-
\]

\[
-\frac{\Delta}{\Sigma}(d\sigma-f_{+}dr-a\cos\theta d\theta+a\sin^{2}\theta d\phi)^{2}.
\]
However, although the so constructed foliation is certainly well-defined
from a mathematical point of view, it does in fact not possess the
conducive property that also $n^{a}$ is locally hypersurface orthogonal
to a fixed representative of associated $r=const.$-hypersurfaces.
However, this shortcoming can be overcome by delivering a DNFS of
Kerr spacetime that is compatible with the given construction of a
NFS. Luckily, precisely such a DNFS has already been provided by Hayward
for Kerr black holes in \cite{hayward2004kerr}, whose extension to
Kerr-Newman seems to be straightforward.

This construction starts, contrary to previous considerations of this
work, in Boyer-Lindquist coordinates. These coordinates, in which
the line element of Kerr spacetime takes the form 
\[
ds^{2}=-dt^{2}+\Sigma(\frac{dr^{2}}{\Delta}+d\theta^{2})+(r^{2}+a^{2})\sin^{2}\theta d{}^{2}+\frac{2Mr}{\Sigma}(dt-a\cdot\sin^{2}\theta d\varphi)^{2},
\]
can be obtained directly from the intially given Kerr coordinate system
by considering the transformations $t=v-r-\frac{2r_{+}}{r_{+}-r_{-}}\ln\vert r-r_{+}\vert-\frac{2r_{-}}{r_{+}-r_{-}}\ln\vert r-r_{-}\vert$
and $\varphi=\phi-\frac{a}{r_{+}-r_{-}}\ln\vert\frac{r-r_{+}}{r-r_{-}}\vert.$ 

Given this setting, the first step in Hayward's construction is to
introduce new coordinates via $(t,r,\theta,\varphi)\rightarrow(t^{*},r^{*},\theta,\varphi^{*})$,
where $t^{*}=t-a\sin\theta$, $r^{*}=\int\frac{\Lambda}{\Delta}dr$,
$\theta=\theta$, $\varphi^{*}=\varphi-\omega_{+}(t-a\sin\theta)$,
where $\omega_{+}=\frac{a}{r_{+}^{2}+a^{2}}$. This is completed in
the second step by another coordinate transformation of the form $(t^{*},r^{*},\theta,\varphi^{*})\rightarrow(X_{+},X_{-},\theta,\varphi^{*})$,
where now $X_{\pm}=\pm e^{\kappa(r^{*}\pm t^{*})}$. This gives the
line element
\[
ds^{2}=-2e^{-m}dX_{+}dX_{-}+q_{AB}(d\theta^{A}+s_{+}^{A}dX_{+}+s_{-}^{A}dX_{-})(d\theta^{B}+s_{+}^{B}dX_{+}+s_{-}^{B}dX_{-}),
\]
whereas, in the given Kerr case, one has $e^{-m}=\frac{\Delta\Sigma}{2\kappa^{2}\Lambda^{2}}e^{-2\kappa r^{*}}$,
$s_{\pm}^{a}=\pm\frac{\Delta}{2\kappa\Lambda^{2}X_{\pm}}(\alpha\partial_{\varphi^{*}}^{a}-a\cos\theta\partial_{\theta}^{a})$
with $\alpha=\omega_{+}(\frac{(r+r_{+})(r^{2}+a^{2})}{r-r_{-}}+r_{+}^{2})$
and $q_{ab}=\frac{1}{\Sigma}((r^{4}+r(r+2M)a^{2}\cos\theta d\theta^{2})d\theta_{a}d\theta_{b}-4Ma^{2}r\cos\theta\sin^{2}\theta d\theta_{(a}d\varphi_{b)}^{*}+\Pi\sin^{2}\theta d\varphi_{a}^{*}d\varphi_{b}^{*})$.
In this new coordinate setting, there holds $r=r(X_{+}X_{-})$ for
the radial function, which has to be determined implicitly with respect
to the relation $X_{+}X_{-}=-e^{2\kappa r^{*}}$.

Adopting these results, it is straightforward to verify that the null
vector fields
\[
\ell_{a}=-e^{-m_{0}}dX_{+a},\;\ell^{a}=e^{m-m_{0}}(\partial_{-}^{a}-s_{-}^{a}),\;n_{a}=-e^{m_{0}-m}dX_{-a},\;n^{a}=e^{m_{_{0}}}(\partial_{+}^{a}-s_{+}^{a})
\]
define a NFS with the desired properties for the choice $e^{-m_{0}}=C\frac{e^{-m}}{\Sigma}\vert_{X_{-}=X_{-}^{0}=const.}$,
where $C=C(X_{+})$.\textcolor{black}{{} The continuous rescaling freedom
of the null normal leads once again to the exterior equivalence class}
$[\ell]$, which encodes the fact that there is a congruence of integral
curves which foliates Kerr spacetime in the considered coordinates.
The associated integral curves are of the form

\begin{align*}
\dot{X}_{+} & =0,\\
\dot{X}_{-} & =e^{m-m_{0}},\\
\dot{\theta}= & e^{m-m_{0}}a\cos\theta\frac{\Delta}{2\kappa\Lambda^{2}X_{+}},\\
\dot{\varphi}^{*}= & -e^{m-m_{0}}\frac{\Delta}{2\kappa\Lambda^{2}X_{+}}\alpha.
\end{align*}
Given these curves, it is straightforward to check that the resulting
NFS leads to the same $2+2$-splitting of the metric as Hayward's
DNFS, but it is in turn much less straightforward to check whether
or not the line element can be expressed in null Gaussian coordinates
and thereby be foliated by a distinct NFS. 

Luckily, the results of a work by Fletcher and Lun \cite{fletcher2003kerr}
seem to indicate that the metrics of both Kerr and Kerr-Newman spacetime
can actually be transformed to so-called generalized Bondi coordinates,
which are special null Gaussian coordinates that have the amazing
property of being specifically adapted to the local geometry of spacetime
in the vicinity of past and future null infinity. In this sense, it
seems to be indeed the case that null Gaussian coordiantes can be
specified with respect to the given Kerr geometric setting and that
therefore a transition from the dual null to the null geometric framework
can be achieved by simply considering a finite number of coordinate
transformations and by simultaneously using different null geodesic
frames for the construction of the respective NFS and DNFS. 

Based on these properties of the models studied, it appears that the
results obtained could possibly bridge some gaps between the dual-null
and standard null approaches to black hole physics. 
\begin{description}
\item [{Acknowledgements:}]~
\end{description}
I want to thank Herbert Balasin for stimulating discussions on the
subject.

\bibliographystyle{plain}
\bibliography{2C__Arbeiten_litfol}

\begin{thebibliography}{10}

\bibitem{arnowitt1959dynamical}
{Richard} Arnowitt, {Stanley} {Deser}, and {Charles}~{W} {Misner}.
\newblock Dynamical structure and definition of energy in general relativity.
\newblock {\em Physical {Review}}, 116(5):1322, 1959.

\bibitem{arnowitt1960canonical}
{Richard} Arnowitt, {Stanley} {Deser}, and {Charles}~{W} {Misner}.
\newblock Canonical variables for general relativity.
\newblock {\em Physical {Review}}, 117(6):1595, 1960.

\bibitem{arnowitt2008republication}
{Richard} Arnowitt, {Stanley} {Deser}, and {Charles}~{W} {Misner}.
\newblock Republication of: {The} dynamics of general relativity.
\newblock {\em General {Relativity} and {Gravitation}}, 40(9):1997--2027, 2008.

\bibitem{ashtekar1991lectures}
{Abhay} Ashtekar.
\newblock {\em Lectures on non-perturbative canonical gravity}, volume~4.
\newblock World {Scientific}, 1991.

\bibitem{brady1996covariant}
{P}{R} {Brady}, {S} {Droz}, {W} {Israel}, and {S}{M} {Morsink}.
\newblock Covariant double-null dynamics: 2+ 2-splitting of the {Einstein}
  equations.
\newblock {\em {Classical} and {Quantum} {Gravity}}, 13(8):2211, 1996.

\bibitem{brown1993quasilocal}
{J}~{David} Brown and {James}~{W} {York}~{Jr}.
\newblock Quasilocal energy and conserved charges derived from the
  gravitational action.
\newblock {\em Physical {Review} {D}}, 47(4):1407, 1993.

\bibitem{chrusciel2012einstein}
{Piotr}~{T.} Chrusciel and {Helmut} {Friedrich}.
\newblock {\em The {Einstein} {Equations} and the {Large} {Scale} {Behavior} of
  {Gravitational} {Fields}: 50 {Years} of the {Cauchy} {Problem} in {General}
  {Relativity}}.
\newblock Birkh{\"a}user, 2012.

\bibitem{d1980covariant}
RA~d'Inverno and J~Smallwood.
\newblock Covariant 2+ 2 formulation of the initial-value problem in general
  relativity.
\newblock {\em Physical Review D}, 22(6):1233, 1980.

\bibitem{fletcher2003kerr}
{Stephen}~{John} Fletcher and {Anthony} {Wah}-{Cheung} {Lun}.
\newblock The {Kerr} spacetime in generalized {Bondi}-{Sachs} coordinates.
\newblock {\em Classical and {Quantum} {Gravity}}, 20(19):4153, 2003.

\bibitem{friedrich1981asymptotic}
{Helmut} {Friedrich}.
\newblock The asymptotic characteristic initial value problem for {Einstein's}
  vacuum field equations as an initial value problem for a first-order
  quasilinear symmetric hyperbolic system.
\newblock In {\em Proceedings of the {Royal} {Society} of {London} {A}:
  {Mathematical}, {Physical} and {Engineering} {Sciences}}, volume 378, pages
  401--421. {The} {Royal} {Society}, 1981.

\bibitem{friedrich1999rigidity}
{Helmut} {Friedrich}, {Istvan} {Racz}, and {Robert}~{M} {Wald}.
\newblock On the rigidity theorem for spacetimes with a stationary event
  horizon or a compact cauchy horizon.
\newblock {\em {Communications} in mathematical physics}, 204(3):691--707,
  1999.

\bibitem{friedrich1983characteristic}
{Helmut} {Friedrich} and {John}~{M} {Stewart}.
\newblock Characteristic initial data and wavefront singularities in general
  relativity.
\newblock In {\em Proceedings of the {Royal} {Society} of {London} {A}:
  Mathematical, Physical and Engineering Sciences}, volume 385, pages 345--371.
  {The} {Royal} {Society}, 1983.

\bibitem{hawking1968gravitational}
{Stephen}~{W} Hawking.
\newblock Gravitational radiation in an expanding universe.
\newblock {\em Journal of {Mathematical} {Physics}}, 9(4):598--604, 1968.

\bibitem{hawking1996gravitational}
{Stephen}~{W} Hawking and {Gary}~{T} {Horowitz}.
\newblock The gravitational {Hamiltonian}, action, entropy and surface terms.
\newblock {\em Classical and {Quantum} {Gravity}}, 13(6):1487, 1996.

\bibitem{hayward1993dual}
{Sean}~{A} Hayward.
\newblock Dual-null dynamics of the {Einstein} field.
\newblock {\em Classical and {Quantum} {Gravity}}, 10(4):779, 1993.

\bibitem{hayward1994general}
{Sean}~{A} {Hayward}.
\newblock {General} laws of black-hole dynamics.
\newblock {\em {Physical} {Review} {D}}, 49(12):6467, 1994.

\bibitem{hayward1994quasilocal}
{Sean}~{A} Hayward.
\newblock Quasilocal gravitational energy.
\newblock {\em Physical {Review} {D}}, 49(2):831, 1994.

\bibitem{hayward2004kerr}
{Sean}~{A} Hayward.
\newblock Kerr black holes in horizon-generating form.
\newblock {\em Physical leview letters}, 92(19):191101, 2004.

\bibitem{isaacson1968harmonic}
{Richard}~{A} {Isaacson} and {Jeffrey} {Winicour}.
\newblock Harmonic and {Null} {Descriptions} of {Gravitational} {Radiation}.
\newblock {\em {Physical} {Review}}, 168(5):1451, 1968.

\bibitem{moncrief1983symmetries}
{Vincent} Moncrief and {James} {Isenberg}.
\newblock Symmetries of cosmological {Cauchy} horizons.
\newblock {\em Communications in {Mathematical} {Physics}}, 89(3):387--413,
  1983.

\bibitem{pawlowski2004spacetimes}
{Tomasz} Pawlowski, {Jerzy} {Lewandowski}, and {Jacek} {Jezierski}.
\newblock Spacetimes foliated by {Killing} horizons.
\newblock {\em Classical and {Quantum} {Gravity}}, 21(4):1237, 2004.

\bibitem{robinson1962some}
{Isaac} Robinson and {Andrezj} {Trautman}.
\newblock Some spherical gravitational waves in general relativity.
\newblock {\em Proceedings of the {Royal} {Society} of {London}. {Series} {A.}
  {Mathematical} and {Physical} {Sciences}}, 265(1323):463--473, 1962.

\bibitem{rovelli2007quantum}
{Carlo} Rovelli.
\newblock {\em Quantum gravity}.
\newblock Cambridge university press, 2007.

\bibitem{sachs1962characteristic}
{Rainer}~{K} Sachs.
\newblock On the characteristic initial value problem in gravitational theory.
\newblock {\em Journal of {Mathematical} {Physics}}, 3(5):908--914, 1962.

\bibitem{stewart1986characteristic}
{John}~{M} {Stewart}.
\newblock The characteristic initial value problem in general relativity.
\newblock In {\em {Astrophysical} {Radiation} {Hydrodynamics}}, pages 531--568.
  {Springer}, 1986.

\bibitem{stewart1982numerical}
John~M Stewart and Helmut Friedrich.
\newblock Numerical relativity. {I}. the characteristic initial value problem.
\newblock In {\em Proceedings of the Royal Society of London A: Mathematical,
  Physical and Engineering Sciences}, volume 384, pages 427--454. {The} {Royal}
  {Society}, 1982.

\bibitem{thiemann2007modern}
{Thomas} Thiemann.
\newblock {\em Modern canonical quantum general relativity}.
\newblock Cambridge {University} {Press}, 2007.

\end{thebibliography}

\end{document}